\documentclass[sigconf]{acmart}
\AtBeginDocument{%
  }

\setcopyright{acmlicensed}
\copyrightyear{2018}
\acmYear{2018}
\acmDOI{XXXXXXX.XXXXXXX}
\acmConference[Conference acronym 'XX]{Make sure to enter the correct
  conference title from your rights confirmation email}{June 03--05,
  2018}{Woodstock, NY}
\acmISBN{978-1-4503-XXXX-X/2018/06}

\usepackage{enumitem}
\usepackage{multirow}
\usepackage{booktabs}

\begin{document}

\title{HiSAC: Hierarchical Sparse Activation Compression for Ultra-long Sequence Modeling in Recommendation Systems}

\author{Kun Yuan}
\authornote{Both authors contributed equally to this research.}
\affiliation{%
  \institution{Alibaba Group}
  \city{Beijing}
  \country{China}
}
\email{yuankun.yuan@taobao.com}

\author{Junyu Bi}
\authornotemark[1]
\affiliation{%
  \institution{Alibaba Group}
  \city{Beijing}
  \country{China}
}
\email{bijunyu.bjy@taobao.com}

% \author{Junyu Bi}
% \authornotemark[1]
% \affiliation{%
%   \institution{Alibaba Group}
%   \city{Beijing}
%   \country{China}}
% \email{bijunyu.bjy@taobao.com}

\author{Daixuan Cheng}
\authornote{Work done during an internship at Alibaba Group.}
\affiliation{%
  \institution{Renmin University}
  \city{Beijing}
  \country{China}}
\email{daixuancheng6@gmail.com}

\author{Changfa Wu}
\affiliation{%
  \institution{Alibaba Group}
  \city{Beijing}
  \country{China}}
\email{wuchangfa.wcf@taobao.com}

\author{Shuwen Xiao}
\affiliation{%
  \institution{Alibaba Group}
  \city{Beijing}
  \country{China}}
\email{shuwen.xsw@taobao.com}

\author{Binbin Cao}
\affiliation{%
  \institution{Alibaba Group}
  \city{Beijing}
  \country{China}}
\email{simon.cbb@taobao.com}

\author{Jian Wu}
\affiliation{%
  \institution{Alibaba Group}
  \city{Beijing}
  \country{China}}
\email{joshuawu.wujian@taobao.com}

\author{Yuning Jiang}
\affiliation{%
  \institution{Alibaba Group}
  \city{Beijing}
  \country{China}}
\email{mengzhu.jyn@taobao.com}

\renewcommand{\shortauthors}{Kun Yuan et al.}

% ================================================================
% ABSTRACT (5-sentence formula: Farquhar)
% ================================================================
\begin{abstract}
We introduce \textbf{HiSAC} (Hierarchical Sparse Activation Compression), a framework that compresses ultra-long user behavior sequences into compact, personalized interest representations for industrial recommendation.
Modeling histories with thousands of interactions is essential for capturing dynamic user preferences, yet existing compression methods suffer from two fundamental limitations: they impose a fixed number of global interest centers that ignore per-user granularity, and they rely on hard assignment that introduces quantization error and discards long-tail behaviors.
HiSAC addresses both via three tightly integrated components: (1)~multimodal tokenization with Residual Quantized VAE that maps items to a hierarchical discrete codebook, (2)~a hierarchical voting mechanism that sparsely activates user-specific \emph{interest-agents} as adaptive preference centers, and (3)~a Soft-Routing Attention that aggregates all historical signals in a debiased semantic space, weighting contributions by similarity to minimize quantization error and preserve sparse long-tail preferences.
We validate HiSAC on a large-scale industrial dataset (200M users, 30B interactions) and the public Taobao-MM benchmark, where it consistently outperforms both generic compression baselines and state-of-the-art sequence models.
Deployed on Taobao's ``Guess What You Like'' homepage serving hundreds of millions of daily active users, HiSAC delivers a \textbf{1.65\%} CTR uplift, 2.24\% CTCVR improvement, and 2.56\% increase in orders while reducing end-to-end inference latency by over 40\%.
\end{abstract}

% ================================================================
% CCS / KEYWORDS
% ================================================================



\maketitle

\vspace{6pt}

% ================================================================
% 1. INTRODUCTION
% ================================================================
\section{Introduction}

Modern recommendation systems in e-commerce and video streaming model user behavior sequences to capture evolving preferences~\cite{zhou2018deep, pi2020search, chang2023twin}.
For highly active users, these histories can exceed 10,000 interactions.
Directly feeding such sequences into attention-based rankers is computationally prohibitive in real-time serving: the quadratic cost of multi-head attention (MHA) over long sequences drives inference latency far beyond production limits, and noise from irrelevant interactions dilutes attention weights, degrading recommendation quality~\cite{zeng2023transformers, cen2020controllable}.

A practical alternative is to summarize the history via a compact set of \emph{interest centers} and model only these summaries.
However, existing compression methods fall short on two fronts:
\begin{enumerate}[leftmargin=*,nosep]
    \item \textbf{Interest center identification.} Methods like ELASTIC~\cite{deng2024elastic} and PolyEncoder~\cite{humeau2019poly} learn a fixed set of \emph{global} interest embeddings shared across all users, ignoring that the number and granularity of interests vary substantially across individuals.
    \item \textbf{Behavior assignment.} Clustering- and hashing-based grouping (K-Means, LSH~\cite{kitaev2020reformer}) and temporal segmentation~\cite{chai2025longer,liao2025multi} rely on hard assignment of behaviors to groups. Temporal splits disregard semantic proximity; hash- and cluster-based methods misassign semantically related items, producing quantization error that disproportionately erases rare long-tail preferences.
\end{enumerate}

We propose \textbf{HiSAC} (Hierarchical Sparse Activation Compression), a framework that replaces fixed global interest centers with \emph{personalized, sparsely activated interest-agents} and replaces hard behavior assignment with \emph{similarity-based soft routing}.
The key insight is twofold: (i)~discretizing items into a hierarchical semantic codebook enables stable, interpretable interest grouping through tree-based voting, and (ii)~decoupling the \emph{routing space} (where similarity is computed using frozen, debiased multimodal embeddings) from the \emph{aggregation space} (where task-specific ranking embeddings are learned end-to-end) preserves semantic diversity while optimizing for downstream accuracy.

HiSAC operates in three stages: (i)~multimodal RQ-VAE tokenization maps items to a hierarchical discrete codebook, (ii)~hierarchical voting sparsely activates personalized interest-agents on a global semantic tree, and (iii)~Soft-Routing Attention aggregates all historical interactions with similarity-weighted attention in a debiased semantic space, reducing quantization error and preserving long-tail signals. Deployed at scale on Taobao's ``Guess What You Like'' homepage---where vanilla MHA over 10k-length sequences exceeds the serving latency budget---HiSAC with production optimizations reduces per-request latency to 21\,ms and, in a two-week online A/B test with hundreds of millions of users, achieves a \textbf{1.65\%} CTR uplift, 2.24\% CTCVR improvement, and 2.56\% increase in orders over the strongest prior compression baseline.

Our main contributions are:
\begin{itemize}[leftmargin=*]
    \item A \textbf{hierarchical voting mechanism} built on multimodal RQ-VAE tokenization that produces user-specific, sparsely activated interest-agents, adapting the number and granularity of interest centers to each individual.
    \item A \textbf{Soft-Routing Attention} mechanism with decoupled semantic and ranking embedding spaces that minimizes quantization error through similarity-weighted aggregation and inherently preserves long-tail preferences.
    \item \textbf{Industrial-scale validation} on a dataset of 200M users and 30B interactions, with online A/B testing demonstrating consistent business metric improvements (1.65\% CTR, 2.56\% orders) alongside substantial latency reduction.
\end{itemize}

% ================================================================
% 2. RELATED WORK
% ================================================================
\section{Related Work}

We organize related work along three methodological axes.

\subsection{Sequence Modeling in Recommendation}

Early attention-based architectures such as DIN~\cite{zhou2018deep} exploited recent behaviors for short-term interest modeling.
Two-stage paradigms like SIM~\cite{pi2020search} and TWIN~\cite{chang2023twin,si2024twin} retrieve a candidate-relevant subsequence before applying attention, trading retrieval--ranking consistency for efficiency.
Transformer-based models~\cite{sun2019bert4rec} and state-space approaches~\cite{liu2024mamba4rec} have been adapted for sequential recommendation, but both incur prohibitive memory costs for sequences exceeding several thousand interactions.
HiSAC takes a different approach: it \emph{compresses} the full history into interest-level representations, preserving global context without per-item computation in the ranking stage.

\subsection{Sequence Compression Strategies}

ELASTIC~\cite{deng2024elastic} and PolyEncoder~\cite{humeau2019poly} compress histories into a fixed set of learnable global embeddings, sacrificing personalization by sharing prototypes across users.
Temporal patching~\cite{chai2025longer,liao2025multi} preserves order but disregards semantic proximity---identical items purchased months apart land in different groups.
LSH~\cite{kitaev2020reformer,datar2004locality} and K-Means~\cite{chong2021k,roy2021efficient} group by feature similarity, but hash collisions and centroid instability introduce variance, and hard assignment produces quantization artifacts.
HiSAC learns \emph{personalized} interest centers via hierarchical voting on a discrete semantic codebook, and uses \emph{soft} assignment to avoid quantization error.

\subsection{Quantized Representation Learning}

VQ-VAE~\cite{van2017neural} introduced discrete codebooks for compact latent representations; VQ-VAE-2~\cite{razavi2019generating} and RQ-VAE~\cite{lee2022autoregressive} extended this with multi-level and residual quantization, improving fidelity.
These methods are well-established in vision and speech, but their application to recommendation has been limited.
HiSAC repurposes the discrete codebook as the backbone for a hierarchical semantic tree, enabling tree-based voting for personalized interest center identification.

% ================================================================
% 3. METHOD
% ================================================================
\section{Method}

\subsection{Problem Formulation and Overview}

Consider a recommendation system with item corpus $\mathcal{I}$ and user set $\mathcal{U}$.
Each user $u \in \mathcal{U}$ has an interaction history ordered by time:
\begin{equation}
    H_u = [h_1, \dots, h_n, \dots, h_N], \quad h_n \in \mathcal{I},
\end{equation}
where $N$ can exceed $10^4$ for highly active users.
Directly modeling $H_u$ with attention-based rankers incurs $\mathcal{O}(N^2)$ or $\mathcal{O}(NM)$ complexity (with $M$ candidates), which is prohibitive in real-time serving.

Our goal is to construct a compressed representation $\hat{H}_u = [\hat{z}_1, \dots, \hat{z}_K]$ where $K \ll N$ and each $\hat{z}_k \in \mathbb{R}^{d'}$ summarizes a distinct aspect of the user's interests.
Unlike retrieval-based truncation, $\hat{H}_u$ is not a subsequence of items but a set of \emph{interest-level} summaries that capture diverse behavioral patterns.

\begin{figure*}[t]
\begin{center}
\includegraphics[width=0.95\textwidth]{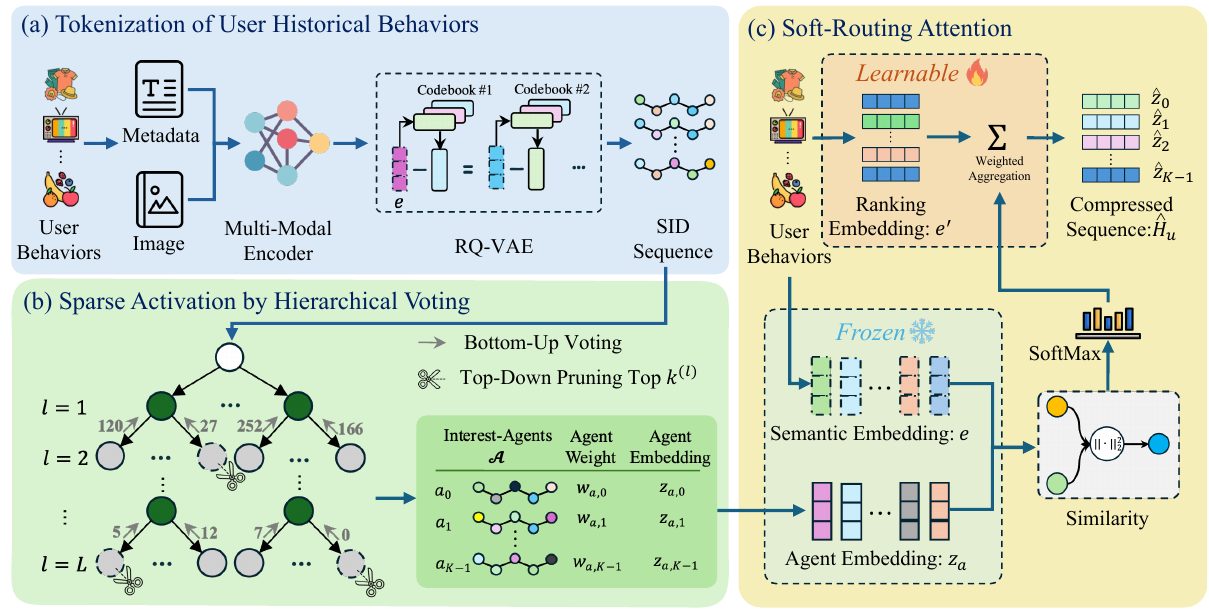}
\end{center}
\caption{Overview of HiSAC. (a)~Items are tokenized into multi-level semantic IDs via a multimodal encoder and RQ-VAE. (b)~A hierarchical voting mechanism projects the tokenized history onto a global semantic tree and prunes layer-wise, yielding personalized interest-agents. (c)~Soft-Routing Attention aggregates all historical interactions for each agent using similarity weights computed from frozen semantic embeddings, with content drawn from trainable ranking embeddings.}
\label{model}
\end{figure*}

Figure~\ref{model} overviews the HiSAC framework, which compresses sequences in three stages:
\begin{enumerate}[leftmargin=*,nosep]
    \item \textbf{Tokenization}~(\S\ref{tokenization}): Items are encoded into $L$-level discrete semantic identifiers (SIDs) via a multimodal encoder and RQ-VAE, producing a global hierarchical codebook.
    \item \textbf{Hierarchical Voting}~(\S\ref{voting}): Each user's tokenized history is aligned to a global semantic tree. Bottom-up vote propagation followed by top-down layer-wise pruning sparsely activates personalized \emph{interest-agents}.
    \item \textbf{Soft-Routing Attention}~(\S\ref{soft-routing}): Each interest agent aggregates all historical interactions via similarity-weighted attention, using decoupled semantic (frozen) and ranking (trainable) embeddings to minimize quantization error and preserve long-tail signals.
\end{enumerate}
Section~\ref{deployment} details the production optimizations that make HiSAC viable under strict latency constraints.

% ================================================================
% 3.2 TOKENIZATION
% ================================================================
\subsection{Tokenization of User Historical Behaviors}\label{tokenization}

In standard recommendation models, item embeddings are learned end-to-end from interaction data, which captures task-specific patterns but inherits biases from item popularity and exposure frequency.
We instead employ a frozen large-scale multimodal encoder to produce semantically rich, debiased item representations.

\textbf{Multimodal Encoding.} For each item $h \in \mathcal{I}$, we collect its descriptive metadata (title, category labels, domain-specific text) and its associated product image.
These signals are processed by a pretrained CLIP ViT-B/32~\cite{radford2021learning} with a Q-Former fusion module~\cite{li2023blip} that applies cross-attention from $n_q{=}8$ learnable query tokens to text and image features, followed by mean pooling to produce a fused embedding $\mathbf{e}_{\text{mm}} \in \mathbb{R}^d$.
This embedding unifies visual and textual semantics in a common space and remains frozen during recommendation training, mitigating exposure bias.
The Q-Former is optimized with a symmetric InfoNCE objective~\cite{he2020momentum} over text and image modalities.

\textbf{Residual Quantization.} Although multimodal embeddings capture rich semantics, they reside in a continuous space where clustering is ambiguous and direct grouping incurs quantization error.
We apply Residual Quantized VAE (RQ-VAE)~\cite{lee2022autoregressive} to discretize $\mathbf{e}_{\text{mm}}$ into an $L$-level hierarchy of codebooks.
RQ-VAE quantizes residuals progressively: the first level quantizes the encoder output; each subsequent level quantizes the residual from the preceding level, yielding increasingly fine-grained partitions.
This produces an $L$-level \emph{semantic identifier} (SID) for each item---a discrete path through the codebook hierarchy that encodes the item's semantics at multiple granularities.
The user's history $H_u$ is correspondingly tokenized into a sequence of SIDs.

\textbf{Why discretization.} The discrete SID representation is the critical enabler of the subsequent stages.
Only with discrete identifiers can we (i)~construct a global semantic tree where nodes correspond to shared codebook entries, (ii)~perform exact vote counting across users by mapping items to deterministic tree nodes, and (iii)~apply layer-wise top-$k$ pruning with guaranteed sparsity.
Continuous embeddings cannot support these operations without expensive nearest-neighbor lookups and unstable cluster assignments.

The RQ-VAE is trained with a reconstruction loss on $\mathbf{e}_{\text{mm}}$ and a commitment loss at each level ($\beta=0.25$), using EMA-based codebook updates with decay 0.99~\cite{razavi2019generating}.

% ================================================================
% 3.3 HIERARCHICAL VOTING
% ================================================================
\subsection{Sparse Activation by Hierarchical Voting}\label{voting}

Given the tokenized history, we construct a personalized set of interest-agents through hierarchical voting.
This addresses our first core challenge: identifying user-specific interest centers at appropriate granularity.

\textbf{Step~1: Global Hierarchical Semantic Tree.}
We organize the entire SID space into a static $(L+1)$-level tree.
Nodes at level $l$ correspond to codeword indices in the $l$-th codebook; the root (level~0) is virtual.
A path from root to leaf uniquely defines a complete SID.
This tree serves as a shared semantic index across all users, enabling efficient hierarchical aggregation without per-user tree construction.

\textbf{Step~2: Voting and Pruning.}
Each item in a user's tokenized history is projected onto its corresponding leaf node.
We then perform two passes:
\begin{enumerate}[leftmargin=*,nosep]
    \item \textbf{Bottom-up vote propagation.} Each leaf receives a vote count equal to the number of items mapped to it. For each internal node, the count is the sum of its descendants' votes.
    \item \textbf{Top-down pruning.} Starting from the root, at each level $l \in \{1,\dots,L\}$, for every activated parent node, we retain only the top-$k^{(l)}$ child nodes by vote count and deactivate the rest.
\end{enumerate}
This layer-wise sparsification focuses computation on the most salient semantic branches.

The remaining active leaf nodes define the user's \textbf{interest-agents}:
\begin{equation}\label{ahent}
    \mathcal{A}_u = \{a_1, \dots, a_K\}, \quad
    \mathcal{W}_u = \{w_{a,1}, \dots, w_{a,K}\},
\end{equation}
where each agent $a$ is represented by its full $L$-level SID path.
The weight $w_a = \text{softmax}_a(\text{count}_a)$ reflects the relative strength of the corresponding interest.
The total number of agents is $K = \prod_{l=1}^L k^{(l)}$, controlled by the per-level budget $\{k^{(l)}\}$.

This adaptive mechanism produces a compact interest summary tuned to each user: users with concentrated interests activate fewer agents (dominated by a few high-vote branches), while those with diverse interests naturally retain more.

% ================================================================
% 3.4 SOFT-ROUTING ATTENTION
% ================================================================
\subsection{Soft-Routing Attention for Compression}\label{soft-routing}

Once interest-agents $\mathcal{A}_u$ are determined, the remaining challenge is to aggregate historical behaviors into each agent's representation.
A naive \textbf{Hard-Routing} approach assigns each interaction to its exact SID-matched agent.
This amplifies quantization error: items whose SIDs are semantically close to but not exactly matching an agent are either assigned to an imperfect agent or discarded.
Long-tail interactions---those whose SIDs do not match any activated agent---are particularly vulnerable.

\textbf{Decoupled embedding spaces.} Our key design choice is to use two distinct embedding spaces for routing and aggregation:
\begin{itemize}[left=0pt]
    \item \textbf{Semantic embedding} $\mathbf{e} \in \mathbb{R}^d$: produced by the frozen multimodal encoder and RQ-VAE (\S\ref{tokenization}), this captures intrinsic cross-modal semantics and is \emph{never updated} during recommendation training. It is used exclusively to compute similarity between items and interest agents, ensuring unbiased routing.
    \item \textbf{Ranking embedding} $\mathbf{e}' \in \mathbb{R}^{d'}$: learned end-to-end via the CTR objective, this encodes interaction-driven signals (user preferences, collaborative patterns) and is concatenated with item-side features (category, brand) and a time-decay feature. It serves as the \emph{content} being aggregated.
\end{itemize}

This decoupling prevents the ranking objective from distorting the semantic space (which would bias routing toward head items) while allowing the aggregated representation to benefit from task-specific optimization.

\textbf{Soft-Routing computation.}
For each interest agent $a$, we reconstruct its semantic prototype $\mathbf{z}_a \in \mathbb{R}^d$ by summing its codewords across all $L$ RQ-VAE codebook levels.
A smaller $L_2$ distance between $\mathbf{z}_a$ and an item's semantic embedding $\mathbf{e}_i$ indicates higher similarity, motivating negative-distance-based attention weights with temperature-controlled softmax:
\begin{equation}\label{eq:atten}
    \hat{\mathbf{z}}_k = w_{a,k}\cdot\sum_{i=1}^{N}
    \underbrace{\frac{\exp\left(-\frac{\|\mathbf{e}_i - \mathbf{z}_{a,k}\|_2}{\tau}\right)}
         {\sum_{j=1}^{N} \exp\left(-\frac{\|\mathbf{e}_j - \mathbf{z}_{a,k}\|_2}{\tau}\right)}}_{\text{semantic attention weight}}
    \cdot \mathbf{e}'_i,
\end{equation}
where $\tau > 0$ controls distribution sharpness and $w_{a,k}$ weights the agent by interest strength.
The final compressed sequence is $\hat{H}_u = \{\hat{\mathbf{z}}_1, \dots, \hat{\mathbf{z}}_K\}$.

This formulation grants each agent access to the \emph{entire} history, with contributions weighted by semantic similarity.
Long-tail items that lack an exact-matching agent can still contribute meaningfully to nearby agents (see Section~\ref{case_study} for a case study).

\textbf{Matrix formulation.}
Let $Z \in \mathbb{R}^{K \times d}$ stack the semantic prototypes of all $K$ agents, $E \in \mathbb{R}^{N \times d}$ stack the semantic embeddings, and $E' \in \mathbb{R}^{N \times d'}$ stack the ranking embeddings of all $N$ historical items.
The pairwise similarity matrix $W \in \mathbb{R}^{K \times N}$ with $W_{i,j} = -\|E_j - Z_i\|_2 / \tau$ yields the compressed sequence via row-wise softmax and element-wise multiplication with agent weights $\mathcal{W}_u$:
$\hat{H}_u = \mathcal{W}_u \odot (\text{Softmax}(W) \cdot E')$.
This mirrors multi-head attention---agents as queries, items as keys and values---with $L_2$ distance replacing dot product for more stable routing in our setting.

% ================================================================
% 3.5 ONLINE DEPLOYMENT
% ================================================================
\subsection{Production Deployment}\label{deployment}

We deploy HiSAC with multi-head attention (MHA)~\cite{vaswani2017attention} for cross modeling between compressed sequences and candidates, with three production optimizations:

\begin{enumerate}[leftmargin=*,nosep]
    \item \textbf{Offline Agent Construction.} After RQ-VAE training, the entire item corpus is tokenized once via the frozen multimodal encoder and codebook to obtain each item's SID; thereafter, only newly added items are processed daily. With SIDs in place, user histories are mapped to the semantic tree, and personalized agents are extracted and stored for online retrieval. Both SID assignment and agent construction are performed offline daily, incurring no additional online cost.
    \item \textbf{Request-Level Compression.} MHA between $N$ tokens and $M$ candidates costs $\mathcal{O}(NM)$. Since the user sequence is identical for all candidates within a request, we compress $N \to K$ once per request ($\mathcal{O}(NK)$), then cross-attend with $M$ candidates ($\mathcal{O}(KM)$), yielding a large reduction when $K \ll N, M$.
    \item \textbf{Cache-Enhanced MHA.} Over 70\% of serving FLOPs come from Q/K/V projections. Candidate queries are pre-computed offline, refreshed every 2 hours; user keys/values are computed once post-compression per request and reused. This preserves exact MHA outputs while cutting projection FLOPs by $\approx$72\% and per-request latency by 38\% (cache hit rate $>$99\%).
\end{enumerate}

\begin{figure}[t]
\begin{center}
\includegraphics[width=0.45\textwidth]{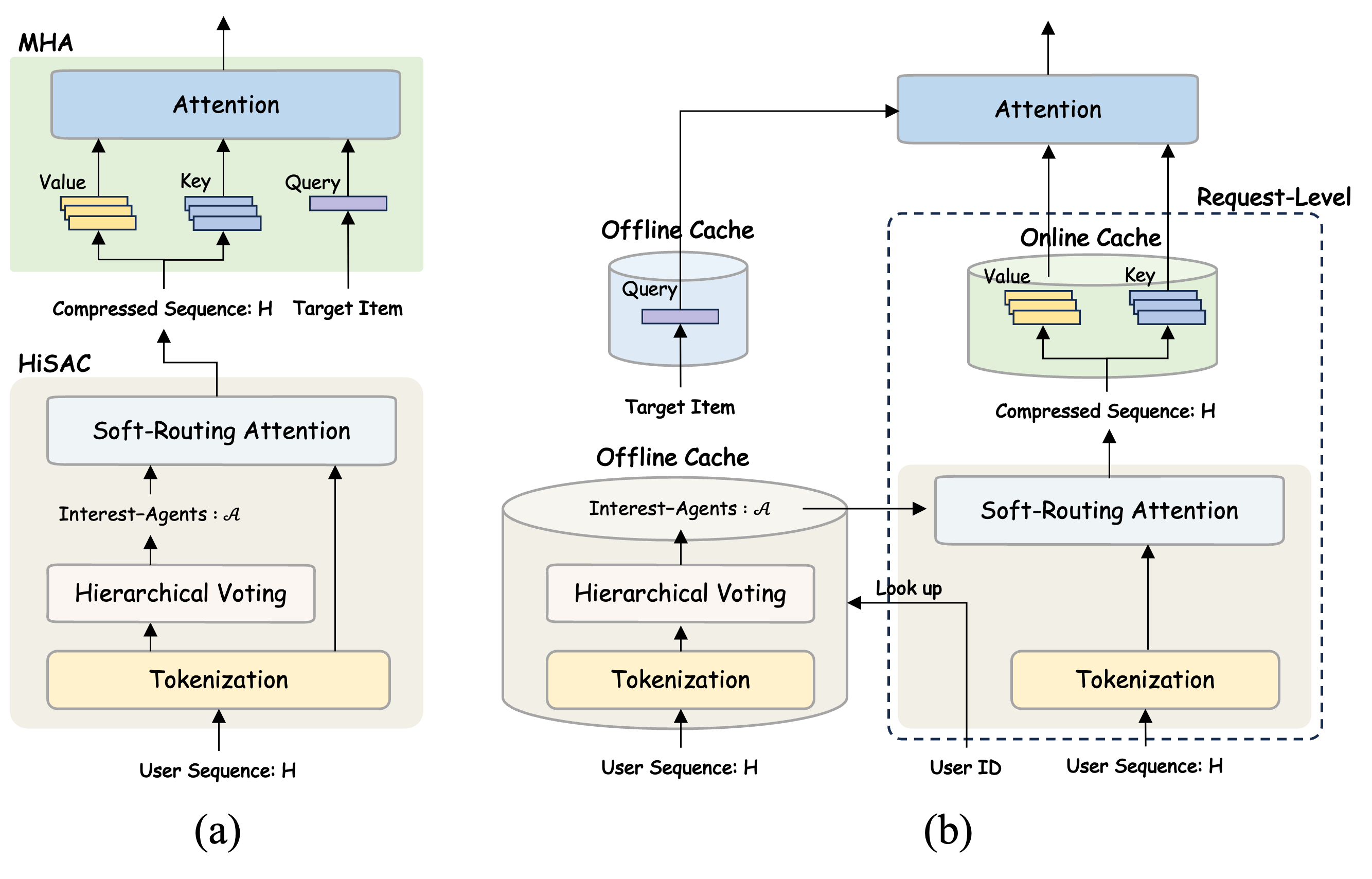}
\end{center}
\caption{Deployment architectures. (a) Standard HiSAC with MHA. (b) Optimized deployment incorporating offline agent construction, request-level compression, and cache-enhanced MHA.}
\label{mha}
\end{figure}

Figure~\ref{mha} contrasts the standard HiSAC deployment with the optimized architecture incorporating these three techniques.

Empirically, with 2,000 candidates and 10k-length histories, vanilla MHA exceeds 100\,ms per request (over budget); HiSAC compression alone reduces this to ${\sim}35$\,ms, and the three optimizations together reach ${\sim}21$\,ms (a 40\% reduction over the compressed baseline) with no measurable loss in recommendation quality.

% ================================================================
% 4. EXPERIMENTS
% ================================================================
\section{Experiments}

\begin{table*}[t]
\caption{Performance comparison on the Industrial Dataset (varying sequence lengths),Taobao-MM and KuaiRec. Best results per metric are \textbf{bold}.}
\centering
\scalebox{1.0}{
    \begin{tabular}{l|cc|cc|cc|cc|cc|cc}
    \toprule
    Dataset & \multicolumn{8}{|c|}{\textbf{Industrial Dataset}} & \multicolumn{2}{|c}{\textbf{Taobao-MM}} & \multicolumn{2}{|c}{\textbf{KuaiRec}} \\
    \midrule
    Sequence Length & \multicolumn{2}{|c|}{1k} & \multicolumn{2}{|c|}{2k} & \multicolumn{2}{|c|}{4k} & \multicolumn{2}{|c}{10k} & \multicolumn{2}{|c}{-} & \multicolumn{2}{|c}{-}\\
    \midrule
    Method & AUC & GAUC & AUC & GAUC & AUC & GAUC & AUC & GAUC & AUC & GAUC & AUC & GAUC\\
    \midrule
    MHA              & 0.6415 & 0.5501 & 0.6420 & 0.5507 & 0.6424 & 0.5511 & 0.6419 & 0.5503 & 0.6692 & 0.6096 & 0.6513	& 0.5772 \\
    Patching         & 0.6385 & 0.5466 & 0.6404 & 0.5485 & 0.6417 & 0.5493 & 0.6420 & 0.5508 & 0.6679 & 0.6084 & 0.6469	& 0.5219  \\
    K-Means          & 0.6407 & 0.5486 & 0.6419 & 0.5504 & 0.6421 & 0.5511 & 0.6426 & 0.5515 & 0.6693 & 0.6092 & 0.6507	& 0.5767  \\
    LSH              & 0.6399 & 0.5470 & 0.6410 & 0.5488 & 0.6418 & 0.5509 & 0.6422 & 0.5512 & 0.6685 & 0.6089 & 0.6501	& 0.5755  \\
    Aggregator       & 0.6427 & 0.5513 & 0.6433 & 0.5517 & 0.6436 & 0.5517 & 0.6438 & 0.5518 & 0.6699 & 0.6101 & 0.6514	& 0.5776  \\
    \textbf{HiSAC}    & \textbf{0.6429} & \textbf{0.5514} & \textbf{0.6438} & \textbf{0.5520} & \textbf{0.6440} & \textbf{0.5522} & \textbf{0.6444} & \textbf{0.5525} & \textbf{0.6704} & \textbf{0.6114} & \textbf{0.6519} & \textbf{0.5789} \\
    \midrule
    PatchRec         & 0.6414 & 0.5498 & 0.6429 & 0.5516 & 0.6435 & 0.5518 & 0.6440 & 0.5518 & 0.6687 & 0.6079 & 0.6691 & 0.6113 \\
    HiSAC + PatchRec & 0.6431 & 0.5514 & 0.6441 & 0.5519 & 0.6441 & 0.5524 & 0.6447 & 0.5530 & 0.6695 & 0.6091 & 0.6699 & 0.6128 \\
    \midrule
    ELASTIC          & 0.6422 & 0.5511 & 0.6434 & 0.5518 & 0.6438 & 0.5521 & 0.6442 & 0.5521 & 0.6709 & 0.6117 & 0.6673 & 0.6097 \\
    HiSAC + ELASTIC  & 0.6433 & 0.5516 & 0.6446 & 0.5524 & 0.6448 & 0.5528 & 0.6449 & 0.5531 & 0.6721 & 0.6126 & 0.6684 & 0.6109 \\
    \midrule
    Longer           & 0.6428 & 0.5514 & 0.6437 & 0.5519 & 0.6443 & 0.5525 & 0.6448 & 0.5529 & 0.6714 & 0.6122 & 0.6702 & 0.6120 \\
    HiSAC + Longer   & 0.6435 & 0.5518 & 0.6449 & 0.5530 & 0.6452 & 0.5533 & 0.6455 & 0.5537 & 0.6733 & 0.6135 & 0.6711 & 0.6132 \\
    \bottomrule
    \end{tabular}
}
\label{tab:main}
\end{table*}

\subsection{Experimental Setup}

\textbf{Datasets.}
We use three datasets:
\begin{itemize}[leftmargin=*]
    \item \textbf{Industrial Dataset}: 14 days of interaction logs from Taobao, comprising 200M users, 1B items, and 30B interaction records. The maximum sequence length is capped at 10,000. Data from the first 13 days serves as training, the final day as validation.
    \item \textbf{Taobao-MM}\footnote{\url{https://huggingface.co/datasets/TaoBao-MM/Taobao-MM}}: A public multimodal dataset~\cite{taobao-mm} with 8.79M users and 35.4M items. User features include ID, age, and gender; item features include ID and category. Maximum sequence length is 1,000.
    \item \textbf{KuaiRec}: KuaiRec~\cite{gao2022kuairec} is a real-world dataset from the recommendation logs of Kuaishou\footnote{\url{www.kuaishou.com}}, containing 7,176 users and 10,728 items. %Unlike the above, KuaiRec provides no multimodal features.
\end{itemize}

\textbf{Metrics.}
Offline: AUC (area under the ROC curve) and GAUC (per-user averaged AUC, which mitigates the influence of highly active users).
Online: Item Page View (IPV), Click-Through Rate (CTR), Click-Through Conversion Rate (CTCVR), and Orders.

\textbf{Baselines.}
We compare against five compression methods combined with MHA:
\begin{itemize}[leftmargin=*]
    \item \textbf{Patching}: groups items by timestamp, preserving temporal order.
    \item \textbf{K-Means}~\cite{chong2021k,roy2021efficient}: clusters tokens by feature similarity with intra-cluster attention.
    \item \textbf{LSH}~\cite{datar2004locality,kitaev2020reformer}: buckets similar tokens via locality-sensitive hashing.
    \item \textbf{Aggregator}: learns a set of global compression tokens, following a similar design philosophy to ELASTIC~\cite{deng2024elastic} and PolyEncoder~\cite{humeau2019poly}.
\end{itemize}
We also integrate HiSAC into three state-of-the-art architectures, replacing their native compression modules:
\begin{itemize}[leftmargin=*]
    \item \textbf{PatchRec}~\cite{liao2025multi}: temporal block-based pooling with multi-scale long-term/short-term modeling.
    \item \textbf{ELASTIC}~\cite{deng2024elastic}: cross-attention to a fixed set of learned representation vectors.
    \item \textbf{Longer}~\cite{chai2025longer}: partitions sequences chronologically, compresses each segment via Linear Attention~\cite{shen2021efficient}.
\end{itemize}
\textbf{Training.}
All models use Adam for dense parameters and Adagrad for sparse parameters, with initial learning rate $1\times10^{-3}$, and all compression methods produce 200 groups.
For the Industrial Dataset, items are encoded by frozen CLIP ViT-B/32~\cite{radford2021learning}; RQ-VAE uses $L=2$ levels of 512 entries each.
For Taobao-MM, we use SCL-based multimodal embeddings~\cite{sheng2024enhancing}; for KuaiRec, which lacks multimodal information, we use the recommendation model's task embeddings directly.
Both use RQ-VAE configured at $200 \times 2$.
Training is conducted on NVIDIA H20 GPUs.
All reported offline metrics are averaged over 3 runs with different random seeds; standard deviation is below 0.0003 for all AUC and GAUC results.

% ================================================================
% 4.2 OVERALL PERFORMANCE
% ================================================================
\subsection{Overall Performance}

Results are in Table~\ref{tab:main}.
\textbf{Compression method comparison.} Patching performs worst, ignoring semantic relationships.
K-Means improves via explicit similarity but incurs high iterative cost and quantization artifacts.
LSH reduces complexity but hash randomness introduces variance.
The Aggregator yields further gains through learned shared representations but sacrifices personalization via fixed global prototypes.
HiSAC achieves the best performance across all settings, with gains widening as sequences grow---consistent with personalized interest-agents and soft-routing best leveraging long-range context.

\textbf{Integration with SOTA architectures.}
Replacing the native compression module of PatchRec, ELASTIC, and Longer with HiSAC consistently improves both AUC and GAUC, demonstrating that HiSAC's semantically coherent groupings generalize across architectures.

% ================================================================
% 4.3 SEQUENCE LENGTH SCALING
% ================================================================
\subsection{Sequence Length Scaling}

To assess how each method leverages increasing amounts of user history, we evaluate AUC across sequence lengths from 1k to beyond 10k.

\begin{figure}[t]
\begin{center}
\includegraphics[width=0.45\textwidth]{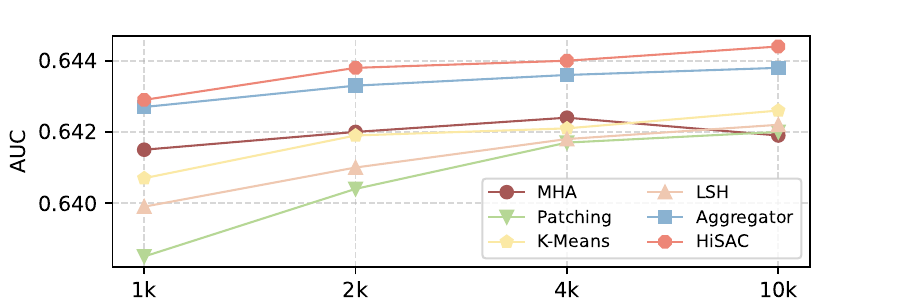}
\end{center}
\caption{Effect of sequence length on AUC. Uncompressed MHA degrades beyond 10k due to attention dilution; compression-based methods improve monotonically. HiSAC achieves the largest gain without saturation.}
\label{length}
\end{figure}

Figure~\ref{length} shows AUC vs.\ sequence length.
Uncompressed MHA declines beyond 10k---attention dilution from noise overwhelming informative signals~\cite{zeng2023transformers,cen2020controllable}.
All compression methods improve monotonically with longer sequences, confirming noise suppression mitigates dilution.
HiSAC achieves the steepest gains without saturation; Patching plateaus earliest.
Aggregator benefits from compression but is constrained by fixed global prototype capacity.

% ================================================================
% 4.4 ABLATION STUDY
% ================================================================
\subsection{Ablation Study}

We isolate the contribution of each core component---multimodal encoding, RQ-VAE, soft-routing attention, long-tail preservation, hierarchical voting, and embedding decoupling---through ablations on the Industrial Dataset with sequence length fixed at 10,000.
Table~\ref{tab:ablation} reports all variants; we analyze them in three thematic groups.

\begin{table}[t]
\caption{Ablation results on the Industrial Dataset (sequence length 10k). All variants use identical training configurations; the ablated component is replaced or removed while others remain unchanged. ``p.p.'' denotes percentage points.}
\centering
\scalebox{0.95}{
\begin{tabular}{l|c|c}
\toprule
Variant                                    & AUC    & GAUC   \\
\midrule
\multicolumn{3}{c}{\textit{Tokenization ablations}} \\
\midrule
w/o Multimodal Encoder (learned embeddings) & 0.6424 & 0.5511 \\
w/o Text Information (image only)          & 0.6429 & 0.5516 \\
w/o Product Image (text only)              & 0.6431 & 0.5517 \\
w/o RQ-VAE (RQ-KMeans instead)             & 0.6438 & 0.5518 \\
\midrule
\multicolumn{3}{c}{\textit{Compression ablations}} \\
\midrule
w/o Soft-Routing Attention (hard routing)  & 0.6429 & 0.5515 \\
w/o Long-Tail Interests (discard unmatched)& 0.6438 & 0.5523 \\
w/o Hierarchical Voting (retain all)       & 0.6445 & 0.5525 \\
\midrule
\multicolumn{3}{c}{\textit{Embedding space design}} \\
\midrule
Both Semantic (single embedding)           & 0.6253 & 0.5392 \\
Aligned (auxiliary loss)                   & 0.6365 & 0.5463 \\
\midrule
\textbf{HiSAC (full)}                      & \textbf{0.6444} & \textbf{0.5525} \\
\bottomrule
\end{tabular}
}
\label{tab:ablation}
\end{table}

\textbf{Tokenization components.}
Removing the multimodal encoder drops AUC by 0.20\,p.p., confirming semantically enriched representations are essential.
Within the encoder, ablating text (--0.15\,p.p.) hurts more than images (--0.13\,p.p.), reflecting text's richer discriminability in e-commerce.
Replacing RQ-VAE with RQ-KMeans reduces AUC by 0.06\,p.p.; RQ-VAE achieves higher Silhouette Coefficient (0.42 vs.\ 0.28) and lower reconstruction error (0.018 vs.\ 0.027), indicating superior cluster separability.

\textbf{Compression mechanisms.}
Replacing Soft-Routing with Hard-Routing drops overall AUC by 0.15\,p.p.
\textit{w/o Long-Tail Interests} recovers 0.09\,p.p., isolating soft assignment's benefit; restoring long-tail interactions adds 0.06\,p.p.
On long-tail items specifically (those outside the top 10\% by exposure), soft-routing improves AUC by \textbf{18.4\%} over hard-routing, confirming that similarity-weighted aggregation captures sparse, niche preferences that hard assignment misses.
Removing hierarchical voting yields a negligible 0.01\,p.p.\ gain while tripling agent count per user ($\sim$3$\times$), proportionally raising online cost---voting distills dominant interests without sacrificing accuracy.

\textbf{Embedding space design.}
Using the same embeddings for routing and aggregation drops AUC by 1.5\,p.p., confirming that semantic representations alone lack the interaction-driven signals needed for effective ranking.
Aligning the two spaces via an auxiliary loss~\cite{he2025plum} partially closes this gap but reduces activated agents per user by 13\% (182$\to$158), indicating that co-optimization steers codebooks toward dominant patterns and narrows interest coverage.
HiSAC therefore keeps the two spaces fully decoupled: frozen semantic embeddings determine routing weights, while trainable ranking embeddings serve as aggregation content. This separation preserves broad interest coverage while optimizing for ranking accuracy.

% ================================================================
% 4.5 CASE STUDY
% ================================================================
\subsection{Qualitative Case Study}

To complement the quantitative ablations, we present a concrete example of how soft-routing weights are assigned within a single user's history.

\begin{figure}[t]
\begin{center}
\includegraphics[width=0.45\textwidth]{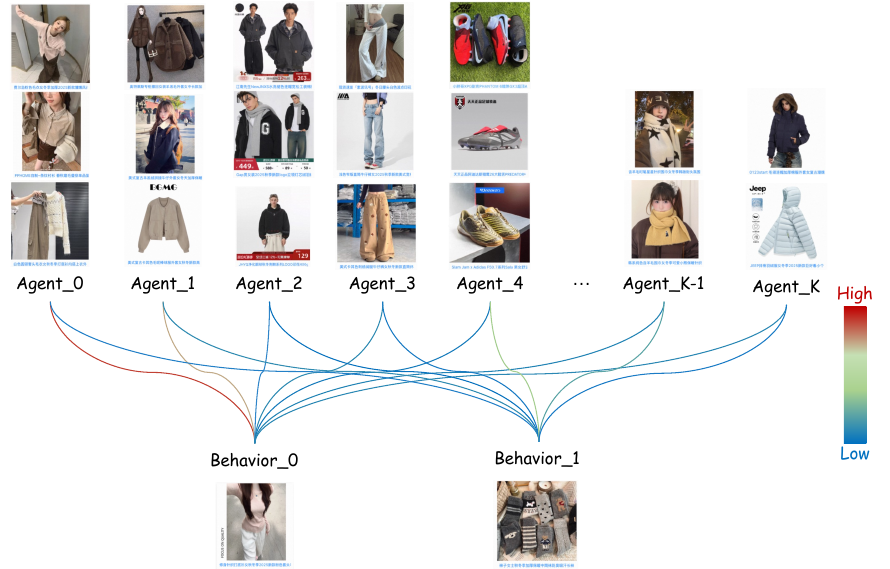}
\end{center}
\caption{Soft-routing example. Behavior\_0 (T-shirt) concentrates weight on Agent\_0 (``topwear''); Behavior\_1 (cotton socks), lacking an exact-match agent, is distributed across semantically proximate agents such as Agent\_4 (``footwear'').}
\label{case_study}
\end{figure}

Figure~\ref{case_study} illustrates this mechanism on a concrete user example, showing how routing weights are assigned to two behaviors from the same user. The dominant behavior (T-shirt) maps cleanly to a dedicated agent and receives concentrated weight. The long-tail behavior (cotton socks), however, has no exact-match agent in the codebook. Under hard-routing, this item would be either misassigned or discarded entirely. Soft-routing instead distributes its contribution across semantically proximate agents---here, Agent\_4 (``footwear'') receives meaningful weight---by exploiting the frozen multimodal embedding space to measure cross-category similarity. This example demonstrates that long-tail preservation in HiSAC is not incidental: the decoupled semantic space provides a stable metric for soft assignment, ensuring that even niche interactions contribute to user representation, consistent with the quantitative gains reported in the ablation study.

% ================================================================
% 4.6 HYPERPARAMETER ANALYSIS
% ================================================================
\subsection{Hyperparameter Analysis}

We study four key design choices under controlled conditions (Industrial Dataset, 10k sequences, $L=2$ codebook levels of 512 entries each, $K \approx 200$ agents unless varied).

\begin{figure}[t]
\begin{center}
\includegraphics[width=0.45\textwidth]{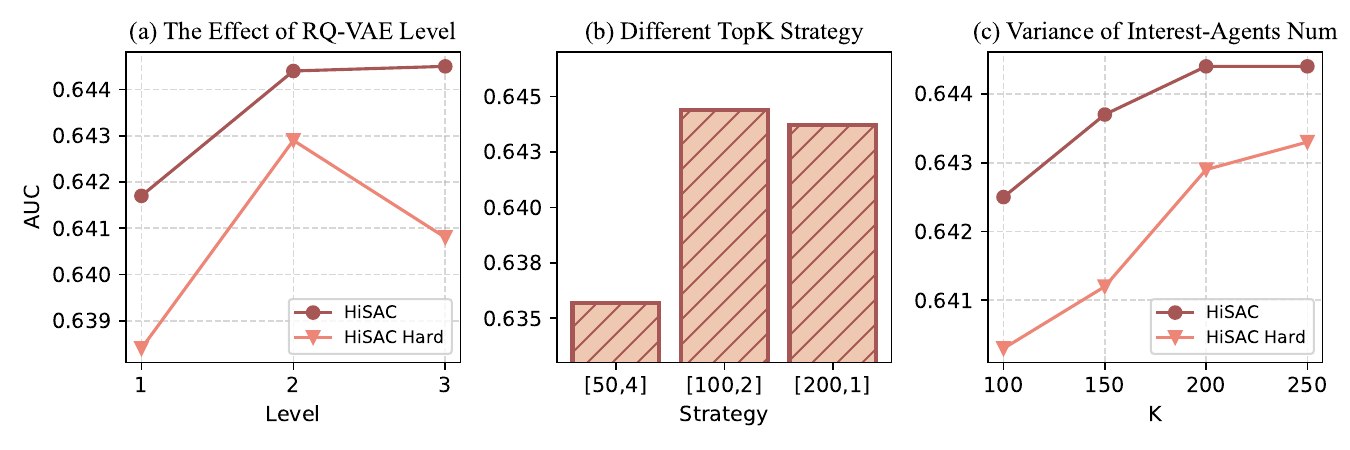}
\end{center}
\caption{Hyperparameter analysis. (a)~Effect of RQ-VAE depth $L$ with controlled agent capacity ($K\approx200$). HiSAC-Hard degrades at $L=3$ due to quantization mismatch; HiSAC remains stable via soft-routing. (b)~Budget allocation across hierarchy levels under total $K=200$; broader coarse-level coverage consistently improves performance. (c)~Effect of total agent budget $K$ under optimal allocation; HiSAC saturates earlier than HiSAC-Hard, confirming superior sample efficiency.}
\label{param}
\end{figure}

\textbf{RQ-VAE depth (Figure~\ref{param}-a).}
Increasing $L$ from 1 to 2 improves performance for both HiSAC and HiSAC-Hard.
At $L=3$, HiSAC plateaus while HiSAC-Hard degrades despite comparable agent counts---soft-routing mitigates the increased quantization mismatch at finer granularity.

\textbf{Hierarchical budget allocation (Figure~\ref{param}-b).}
With $K=200$, $L=2$, comparing $(k^{(1)}, k^{(2)}) = (50,4), (100,2), (200,1)$: performance improves as budget shifts to the coarser level, since per-category user interests concentrate on few fine-grained subcategories.

\textbf{Total agent budget (Figure~\ref{param}-c).}
HiSAC saturates at $K \approx 200$; HiSAC-Hard improves through $K=250$, confirming soft-routing's superior sample efficiency---it exploits long-tail interactions without exhaustive exact-match agent coverage.

\begin{figure}[t]
\begin{center}
\includegraphics[width=0.45\textwidth]{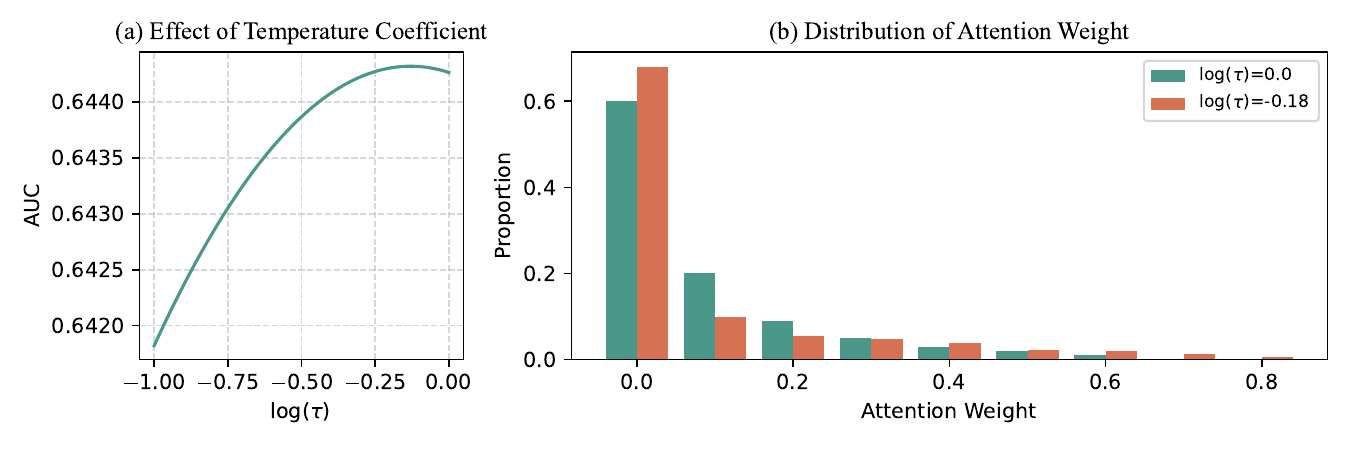}
\end{center}
\caption{Impact of temperature $\tau$. (a)~AUC vs.\ $\log(\tau)$, peaking at $\log(\tau) = -0.18$. (b)~Distribution of attention weights at different $\tau$ values. Lower $\tau$ sharpens the distribution, concentrating weight on the most similar agents; excessively small $\tau$ degenerates into near-hard routing and loses the benefit of soft aggregation.}
\label{param:tau}
\end{figure}

\textbf{Temperature $\tau$ (Figure~\ref{param:tau}).}
AUC peaks at $\log(\tau) = -0.18$. Smaller $\tau$ sharpens attention, suppressing noise; excessively small values collapse into near-hard routing. The optimum balances selectivity with inclusiveness.

% ================================================================
% 4.7 ONLINE A/B TEST
% ================================================================
\subsection{Online A/B Testing}

To verify that offline improvements carry over to real-world serving, we conducted a controlled two-week A/B test on Taobao's ``Guess What You Like'' homepage, serving hundreds of millions of daily active users, against the strongest prior compression method in production.

\begin{table}[t]
\caption{Online A/B test results. All improvements are statistically significant.}
\centering
\scalebox{1.0}{
\begin{tabular}{c|cccc}
\toprule
Method & IPV       & CTR       & CTCVR     & Orders    \\
\midrule
HiSAC  & +1.93\%   & +1.65\%   & +2.24\%   & +2.56\%   \\
\bottomrule
\end{tabular}
}
\label{tab:abtest}
\end{table}

HiSAC improves all metrics (Table~\ref{tab:abtest}): CTR increases 1.65\% and IPV 1.93\%; CTCVR and orders improve 2.24\% and 2.56\%, indicating engagements translate into purchases.

We further examine soft-routing's effect on long-tail items. In production, the top 10\% most frequent items capture over 90\% of exposure, skewing predictions toward head content. Under HiSAC, the CTR of items outside this top 10\% improves by \textbf{12.6\%}, and the average number of distinct item categories surfaced per user increases by \textbf{5.7\%}. These results directly validate soft-routing's ability to preserve underrepresented preference signals that hard-assignment baselines discard.

% ================================================================
% 5. CONCLUSION
% ================================================================
\section{Conclusion}

We presented HiSAC, a hierarchical semantic-agent compression framework that addresses two persistent challenges in ultra-long sequence modeling for recommendation: identifying user-specific interest centers and reliably assigning behaviors to them.
HiSAC combines multimodal RQ-VAE tokenization, hierarchical voting for sparse agent activation, and Soft-Routing Attention with decoupled embedding spaces.
On a large-scale industrial dataset and the public Taobao-MM benchmark, HiSAC consistently outperforms generic compression baselines and state-of-the-art sequence models.
Deployed on Taobao's homepage, it delivers a 1.65\% CTR uplift, 2.56\% orders increase, and over 40\% latency reduction.
These results demonstrate that semantically grounded, personalized compression is both more accurate and more efficient than global-prototype or hard-assignment approaches.

% Future work includes online incremental agent updates, domain-adaptive quantization depth, and extension to multi-scenario recommendation.

% ================================================================
% BIBLIOGRAPHY
% ================================================================
\section*{GenAI Usage Disclosure}

During the preparation of this work, the authors used generative AI tools (ChatGPT and similar large language model assistants) solely to assist with language polishing, including grammar correction, phrasing refinement, and sentence-level clarity improvements, in a manner analogous to using a writing assistant such as Grammarly. These tools were \emph{not} used to generate any technical content, including research ideas, methodology, experimental design, results, figures, tables, theoretical analyses, or code.  All scientific contributions, claims, and interpretations are entirely those of the authors, who reviewed and edited any AI-assisted text and take full responsibility for the content of this paper.
\bibliographystyle{ACM-Reference-Format}
\bibliography{hisac}

\end{document}